\documentstyle[prl,aps,epsf,multicol,amssymb]{revtex}
\begin{document}
\draft
\title{\bf{On the Incommensurate Phase of Pure and
Doped Spin-Peierls System CuGeO$_3$}}
\author{Somendra M. Bhattacharjee $^{a}$, Thomas Nattermann $^{b}$ and
Christopher Ronnewinkel $^{b}$}
\address{$^{a}$ Institute of Physics, Bhubaneswar 751\,005, India\\
$^b$ Institut f\"ur Theoretische Physik,
Universit\"at zu K\"oln, 50937 K\"oln, Germany.\\}
\date{\today}

\maketitle
\begin{abstract}
Phases and phase transitions in pure and doped spin-Peierls 
system CuGeO$_3$ are considered on the basis of a Landau-theory. In 
particular we discuss  the critical behaviour, the soliton width and 
the low temperature specific heat of the incommensurate phase. 
We show, that dilution leads always to the destruction of long range order
in this phase, which is replaced by an algebraic decay of
correlations if the disorder is weak.
\end{abstract}

\parskip0pt
\begin{multicols}{2}
The spin-Peierls (SP) transition  is the classic instability of one
dimensional quantum spin-half antiferromagnetic chains due to the
coupling of the spins with the lattice. A rigid Heisenberg chain has
a nonmagnetic uniform ground state with a gapless fermionic excitation
spectrum\cite{bethe}. This can be seen most easily by using the
Jordan-Wigner transformation,  which maps the spins onto (strongly)
interacting pseudo-fermions\cite{fradkin}. 
Due to the  coupling to the lattice the system can lower its 
energy by the standard Peierls
mechanism: lattice distortions freeze in at a wave vector $2k_F$
which leads simultaneously to the opening of a gap at the Fermi-level 
in the fermionic spectrum such that the energies of all occupied
fermionic states decrease\cite{peierls}. 
In zero magnetic field  the free-fermion band
is half-filled with Fermi-wave vector $k_F=\frac{\pi}{2a}$, which
corresponds to a dimerization of the chain. A non-zero magnetic
field lowers the Fermi-level\cite{fermi}, but Umklapp processes
still favor the distortion at $\pi /a$ until a critical field strength
$H_I$ is reached, at which a transition to an incommensurate (I) phase
with modulation vector $|2k_F-q_s|$ sets in. In the 
I-phase a new (empty) band appears in the middle of the gap of the fermionic
spectrum. Thus, spin excitations still exhibit a gap which is however
smaller than the gap of the dimerized (D) phase.
The above picture follows from theories obtained for free
or weakly interacting pseudo fermions, in which phonon
dynamics were essentially ignored\cite{MFA,RPA}.
There, the SP transition 
is the result of the freezing of a (classical) phonon mode
due to further downwards renormalization of the phonon frequency by 
the spin-phonon interaction\cite{MFA,RPA}. This scenario is supported by the
experimental data of organic SP systems \cite{organic}. However, it
does not seem to apply in all respects to the transition found recently in
the inorganic SP substance CuGeO$_3$ \cite{experiments.review}.
Though not devoid of controversy, there is now a wealth of
well-accepted results for  CuGeO$_3$, which shows two SP transitions 
\cite{neutron,spec.heat,solitons.exp,magnetization,thermal.exp,frustration,doping.exp}.
The SP transition from a disordered, uniform (U) to a D phase
at $14.3{\rm\,K}$ in zero field is shifted slightly to lower $T$
if the field increases until a  Lifshitz point at $T\approx 11.3{\rm\,K}$ and
$H\approx 12.5{\rm\,T}$ is reached, where the transition to an I phase sets in.
Some experimental results which are not explained by the existing
theories, are:
{ (i)} no soft phonon has been observed so
far\cite{experiments.review,neutron},
{ (ii)} a (Peierls) gap
in the D phase is observed in low temperature specific heat
measurements, but not in the I phase, 
for which a Debye--like $T^3$-law has been found with
an amplitude much larger than the background (lattice) 
contribution\cite{spec.heat},
{ (iii)} solitons, which are supposed to produce the modulation in the I
phase are broad \cite{solitons.exp} in comparison to the sharp 
Sine-Gordon like solitons 
predicted by mean-field like calculations\cite{MFA},
{ (iv)} already a small amount of doping leads to a strong reduction of
the SP temperature $T_{\rm SP}$\cite{doping.exp} and a drastic suppression of 
the anomalies at the UI transition\cite{solitons.exp,buechner}.

Since the phonon energies are always large compared with the magnetic
ones, the applicability of the adiabatic approximations has been
questioned \cite{adiabatic}.
Khomskii et al. \cite{khomskii} developed a soliton
picture of the SP transition in CuGeO$_3$, which resembles somewhat 
structural order-disorder transitions \cite{struc.trans}.
No soft mode phonon is expected, but the SP transition corresponds to 
deconfinement of solitons, which are bound to pairs below $T_{\rm SP}$.
These solitons are simultaneously magnetic and structural excitations:
they carry spin $1/2$ and are domain walls between the two
groundstates of the dimerized lattice.

It is the aim of the present paper to explain the properties {(ii)}--{(iv)}
by a pure phenomenological approach, which avoids delicate
approximations in the coupled spin phonon system:
The $T^{3}$-law of the specific heat in
the I phase is explained quantitatively by phason fluctuations. It is
argued that broad
solitons are fingerprints of the type II lock-in
transition which occurs in SP systems like CuGeO$_{3}$.
Finally, we show that dilution leads to complete destruction of long
range order in the I phase. 

Incommensurate phases are classified according to the existence of an
inversion symmetry for the
structural transition in question \cite{bruce}.  In case there is an
inversion symmetry for the Hamiltonian, as for CuGeO$_3$, 
first derivatives of the order parameter (Lifshitz
invariants) do not exist.
Indeed, for
CuGeO$_3$ the uniform high temperature orthorhombic structure, space
group $Pbmm$,
changes below $T_{\rm SP}$ to a dimerized structure,
space group $Bbcm$, with distortion wave vector $(\frac{1}{2}
\frac{1}{2} 0)$ (established from X-ray and neutron diffraction
experiments\cite{neutron}). Standard group theoretic arguments 
based on the symmetries and the invariant group of the distortion
vector in the Brillouin zone\cite{mukamel} show,
that the transition is described by four non-equivalent, one
dimensional irreducible representations\cite{braden}. It is very likely, that only one
of these four representations corresponds to the primary order
parameter, which is real and can be considered to be proportional to
the displacement of the copper ions. The other three may occur as secondary 
order parameters.
In fact, neutron scattering data indicates that at least two normal
modes are necessary to explain the displacement pattern of the D phase\cite{neutron}.
A simple
transformation of reversing the displacements on one sublattice helps
us in getting an ordered state with zero wavevector for the D
phase. We take this transformed and coarse grained displacements to
be proportional to the
order parameter field $\psi({\bf x})$. In general, $\psi({\bf x})$
will also include contributions from the magnetic degrees of freedom.

The Landau Hamiltonian is that of an anisotropic Ising model 
\cite{horn,bruce} 
${\cal H}\{\psi\}= \int d^d {\bf x}\,h\{\psi({\bf x})\}$ ,
where the Hamiltonian density is (for $i=x,y,z$)  
\begin{eqnarray}
\label{ham1}
h\{\psi\} = \frac{r}{2}\psi^2 &+& 
{\sum\nolimits_i}\frac{c_i }{2} 
  (\partial_i \psi)^2 +  \frac{u}{4}\psi^4+ \nonumber\\
&+& \frac{d}{4} (\partial_z^2 \psi)^2 + 
\frac{w}{6}\psi^6 +
\frac{e}{2}\psi^2\left(\partial_z\psi\right)^2
\end{eqnarray}
Terms involving $\partial_z\psi\,\partial_z^3\psi$, though not shown 
explicitly, may also appear\cite{bruce}.
We have included higher order 
terms in order to stabilize the system for the case when 
one of the $c_i$ coefficients 
(here $c_z$) or $u$ becomes negative for 
sufficiently strong magnetic field. A negative $c_i$
signals the transition to the I phase.  
The parameters $r$ and $c_i$ are
taken as analytic functions of $T$ and $H$ with $r=r_0\,(T-T_{\rm SP}(H)) $, 
and $c_z(T,H)=c_0\,(H_{\rm I}(T)-H)$. Microscopic treatments \cite{MFA} and
experiments suggest \cite{spec.heat}, that $u$ also decreases
considerably with increasing magnetic field. 


A {\it mean field analysis} (MFA) of the phase diagram requires treating the
Hamiltonian Eq. (\ref{ham1}) as a free energy
minimized with respect to $\psi$.  Let us first assume, that $u$ and $d$
remain positive everywhere in the $H$--$T$--plane. Then, we can
ignore the last two terms  in (\ref{ham1}).
If $c_{i}>0 $ for all $i$, the
minimum of the free energy occurs for wave vector ${\bf k}=0$, while a nonzero
${\bf k}$ vector is possible if $c_z <0$.  The mean field phase
boundaries are given for UD: $r=0$, $c_z>0$, 
UI: $r=c_z^2/2d$ and DI: $r=\frac{1}{4} (\sqrt{3/2} -1)^{-1}
c_z^2/d$\cite{horn}.  The spontaneous wave vector in the I phase
 is given by $q_s^2=\!-{c_z}/{d}$. The DI transition is
{\it first order} while the other two are second
order, in agreement with experiments.


{\it Fluctuations} do make subtle changes in the phase diagram but the
overall features remain the same.  The critical behavior of the UD
transition is Ising--like but the UI transition is XY--like. 
This difference originates from the fact, that in the I phase the order
parameter condenses at ${\bf k}=\pm {\bf q}_s$ and has consequently two
components ${\bf \Delta}({\bf x})=(\Delta_1({\bf x}),\Delta_2({\bf x}))$. These are related to
$\psi ({\bf x})$ by
$\psi ({\bf x})=\sqrt{2}\left[ \Delta_1({\bf x})\cos({\bf q}_s{\bf x})+
\Delta_2({\bf x})\sin({\bf q}_s{\bf x})\right]$.
In the I phase, the Landau functional thus can be written as 
\begin{eqnarray}
\label {xyfree}
h\{{\bf \Delta}\} = \textstyle\frac{1}{2}(r&-&c_z^2/2d){{\bf \Delta}^2} + 2
|c_z | ({\partial_z {\bf \Delta}})^2 \nonumber\\ 
&+& c_x({\partial_x {\bf \Delta}})^2\!+ c_y({\partial_y {\bf \Delta}})^2
\!+\textstyle\frac{3}{2} u ({\bf \Delta}^2)^2
\end{eqnarray}
Since the number of degrees of freedom of the system cannot change
when going from the D to the I phase, it is clear that Eq. (\ref{xyfree})
is valid only for fluctuations of ${\bf \Delta}({\bf x})$ with  wavelength
long compared to $q^{-1}_s$, i.e. as long as we are away from the Lifshitz
point.

If one approaches the ordered phase along the line
$c_z(T,H)=0$, one observes so called {\it Lifshitz critical behaviour},
which follows from a change of the dispersion relation to
$A_{\bf k}=r+c_xk_x^2+c_yk_y^2+dk_z^4$. 
Note that at the Lifshitz 
critical point the conventional hyperscaling is changed to
$\nu_{\parallel}+(d-1)\nu_{\perp}=2-\alpha$ where 
$\nu_{\parallel}=\nu_{\perp}/2=0.31$ are the correlation length 
exponents parallel and perpendicular to
the $z$-direction\cite {horn}. 
Approaching the D or I phase, respectively,
from the U phase on a line parallel to 
that given by $c_z(T,H)=0$, at first
Lifshitz type critical behaviour will be observed before 
the region of Ising-- or XY--type critical, respectively, 
behaviour is asymptotically reached. 


Considering the DI transition, fluctuation effects are expected to be
less important, because it is first order in MFA. A refined
MFA has been worked out by Bruce, Cowley and Murray \cite{bruce} 
for this case, who 
found that in the I phase the order parameter can be described
by a multiplane-wave Ansatz
$\psi ({\bf x})=\sum  a_m\cos{(mq_zz)}$
with $m=1,3,5...$, which is 
rapidly converging. For example, the ratio 
$|a_3/a_1|\approx 0.035$ close to the transition\cite{bruce}.
In this sense the system shows {\it broad}
domain walls. 
Also in this refined theory the transition remains first order.

Above we assumed $u$ to be {\em positive} even
for large field values.
In the opposite case,  the transition to the U phase
might become first order. Some mean-field theories\cite{MFA} predict 
very special relations between the coefficients of the Landau-expansion, i.e.
\begin{equation}
{u}/{c_z}={\rm const.}\;,\quad w={3du^2}/{4c_z^2}\;,\quad e={5du}/{2c_z}\;.
\label{eq:omega,e}
\end{equation}
If these are fulfilled, the DI transition may 
become {\it continuous}, at least close to the Lifshitz
point\cite{aharony}. Indeed, 
for this very particular relation of the
coefficients of the Landau-expansion (\ref{ham1}), the ground state
solutions are the Jacobian elliptic functions
$\psi_s (z)\equiv \sin(\phi_s(z)/2)=\psi_0\,{\rm sn}({z}/{k\xi_s},k)$,
where $\xi_s $ is a bare correlation length (expressed by $c_z$, $d$ and $r$) 
and $k$ is the modulus of this function \cite{MFA,Zang}. Note, that 
$\phi_s (z)$, which is related to the spin-density, obeys 
the Sine-Gordon equation. In this case solitons are {\it sharp} in
the sense that the separation of domain walls diverges by approaching
the D phase. However, from a symmetry
point of view, which we adopt here, we do not see a deeper reason, 
why the relations (\ref{eq:omega,e}) should be fulfilled in general by an 
{\it exact} microscopic theory. In fact, these relations were obtained 
using the adiabatic
approximation. Consequently one has to expect, that in general the $w$- 
and $e$-terms in (\ref{ham1}) {\it do} exist, but violate 
the relations (\ref{eq:omega,e}). These terms will change the  
modulation amplitude ratio $|a_3/a_1|$ to larger values, but without reaching 
the sharp soliton limit. Thus the DI transition is expected to remain
first order, as found also experimentally  for ${\rm CuGeO}_3$ 
\cite{experiments.review,solitons.exp}.


Although the validity of Landau-theory is essentially restricted to 
the region close to the transition, one should expect that it can be used
to understand at least qualitatively the low temperature specific heat 
data. For this purpose, we have to determine the low-lying excitations 
of the ordered structure.
These can be found by adding the kinetic energy term
$\int {\rm d}^dx\frac{\rho}{2}\dot{\psi}^2({\bf x})$ to the
GL-Hamiltonian, where the mass density $\rho$ 
will have contributions both from the
magnetic and the lattice degrees of freedom. We will further assume,
that $\psi $ obeys Bose-statistics.
Using the saddle point approximation to determine the equilibrium
value of $\psi$, one obtains in the  D phase
$\omega^2({\bf k})=\frac{1}{\rho}(2|r| +c_ik_i^2)$ for the
frequency of the harmonic excitations of the order parameter field.
In the D phase, were the order parameter is real, 
we identify $E_g=\hbar (2|r|/\rho )^{1/2}$ with the gap which is found in
the low-T specific heat.
In the I phase in addition to the massive 
amplitude mode a gapless {\it phason} mode with frequency 
$\omega^2({\bf k})=\frac{1}{\rho}\left( c_xk_x^2+c_yk_y^2+2|c_z|k_z^2\right)$
 appears\cite{bruce}, which will dominate the specific heat 
\begin{equation}
C_{\rm phason}\approx\frac{{\sqrt 2}\pi^2}{15}k_B\left(\frac{k_BT}{E_g
\overline{\xi}_{0}}\right)^3 \equiv \beta _{\rm phason}T^3 .
\label{eq:C-phason}
\end{equation}
Here we have introduced $\overline{\xi}_{0}=(\xi_{0x}\xi_{0y}\xi_{0z})^{1/3}$ where 
$\xi_{0i}^2=c_i/r_0T_0$ and used $T\approx T_0/2$ to express $\rho$ by $E_g$. 
Thus, the phason mode delivers a 
$T^3$--contribution to the low--temperature specific heat, in
addition to that from acoustic phonons\cite{foot}. 
 

Next we  extend our analysis to the
{\it quenched disordered case}, e. g. random substitutions of Cu by Zn or Ni
and/or Ge by Si in CuGeO$_3$. Such substitutions change the various
interactions locally but do not break the symmetry of the
displacements in favor of a particular dimerization.  Therefore, the
effects of these random substitutions can be modeled by randomness in
the coefficients of the original Landau Hamiltonian without
any symmetry breaking term. Little reflection shows, that
the main effect will come from a randomness 
$\delta r({\bf x})$ in $r$ \cite{deltac}.
In the D phase, this leads to a decrease of $T_{\rm SP}$, as was shown  
microscopically by Khomskii et al.\cite{khomskii}. Moreover, the critical 
behaviour will 
be changed to that of the diluted Ising model \cite{dil.Ising}.

The effect of disorder is even more severe in the I phase. This can be 
seen easily by rewriting $\psi ({\bf x})$ as 
$\psi({\bf x}) =  \sqrt{2}\Delta({\bf x})\cos{({\bf q}_s{\bf x}+
\theta ({\bf x}))}$.
With $\delta r({\bf x})=\kappa\sum_i\delta ({\bf x}-{\bf x}_i)$
the disorder term can now be written as 
\begin{equation}
\frac{\kappa}{2}\sum\limits_i\Delta^2({\bf x}_i)\,\cos\left[2\left(\theta ({\bf x}_i)+
{\bf q}_s{\bf x}_i\right)\right]
\label{rand.anis}
\end{equation}
The random impurity positions ${\bf x}_i$ lead to a random phase
$\alpha_i\equiv \alpha ({\bf x}_i)=2{\bf q}_s{\bf x}_i\,({\rm mod}\;2\pi)$ which
 is equally distributed between 0 and $2\pi$.
It is well known that such a random anisotropy term destroys the
translational long range order of the I phase\cite{IM}.
However, the phase--phase correlation function diverges only logarithmically
\cite{Vill.Fern}
$\overline{\langle (\theta ({\bf x})-\theta({\bf 0}))^2\rangle} =\frac{\pi^2}{18}\ln(x/L_L)$.
Here the overbar denotes the disorder average.
The Larkin--length $L_L$  is related to the strength of the
disorder, a rough estimate is 
\begin{equation}
L_L\approx 2\pi^3\left[\bar{\xi}_0^2/({d \ln T_{\rm SP}}/{d n_{\rm imp}})
\right]^2 n_{\rm imp}^{-1}
\label{eq:Larkin}
\end{equation}
where $n_{\rm imp}$ denotes the concentration of the impurities.
Because of the 
logarithmic divergence of the phase fluctuations, there is however 
{\it quasi--long range order} of the order parameter correlation function
\begin{equation}
\overline{\left<{\bf\Delta}({\bf x}){\bf\Delta} ({\bf 0})\right>}
\approx\left(\bar{\xi}_0\sqrt{{\textstyle\sum_{i}}(x_i/\xi_{0,i})^2}/L_L\right)^{-\pi^2/36} 
\label{eq:PP}
\end{equation}
Despite of the  loss of true long range order, the system
will however still show Bragg peaks of finite width, as follows from
the Fourier transform of (\ref{eq:PP}).
In deriving these results we have neglected vortex-ring
excitations. It has been argued recently, that these can indeed be neglected for
sufficiently weak disorder and low temperatures \cite{vortexrings}. At
elevated temperatures or larger dilution their condensation triggers
the transition to the disordered phase. The type of this transition is
presently unknown.


We briefly apply the results obtained so far to ${\rm CuGeO}_3$. 
Fixing the  $T=0$ 
value of the order parameter at $\psi_0=1$, we
have $r_0T_0=u_0$. From the mean--field jump of the specific heat
$\Delta C_{\rm MFA}=r_{0}^{2}T_{0}/2u=u_0/2T_0\approx 22.7\frac{\rm mJ}{\rm K\,cm^3}$
in zero magnetic field\cite{spec.heat}, we get
$u_0=650\frac{\rm mJ}{\rm cm^{3}}$
which gives the correct size of the critical region.
Since $\Delta C_{\rm MFA}$ decreases for increasing field and is reduced 
approximately by a factor $4.6$ when reaching the Lifshitz point, $u$ is 
reduced correspondingly to about  $u_L=112\frac{\rm mJ}{\rm cm^3}$, but still positive.
Defining the Ginzburg critical region 
$\tau_G\equiv |T_G-T_0|/T_0$ as the region, in which 
the first fluctuation correction
to the specific heat becomes larger than $\Delta C_{\rm MFA}$, this
gives for zero field, $\xi_{0,x}=0.12\,\rm nm$, $\xi_{0,y}=0.36\,\rm nm$
and $\xi_{0,z}=0.69\,\rm nm$\cite{corr.lengths}, with geometric mean $\overline{\xi}_0=0.31\,\rm nm$,
$\tau_{G,I}\approx(k_BT_0/8\pi u\smash{\overline{\xi}}_0^3)^{2} \approx 0.16$ --
larger $\Delta C_{\rm MFA}$
diminishes $\tau_G$ correspondingly. 
For the XY-transition far from the Lifshitz point we 
get \hbox{$\tau _{G,XY}\approx 0.32$} at a magnetic field where
\hbox{$\xi_{z}(H, T=T_{\rm SP})\approx\xi_{0,z}$}.
At the Lifshitz point the
critical region is  given by 
$\tau_{G,L}\approx \tau_{\smash{G}}^{2/3}(\xi_{0,z}u_0/{\sqrt 2} \zeta u_L )^{4/3} \approx 0.7$ 
where \hbox{$\zeta_0=(d/2r_0T_0)^{1/4} \approx 1.2\,\rm nm$}. 
The critical exponent
$\beta$ changes from $\beta_I\simeq 0.325$ for $H<H_L$
to $\beta_L\simeq 0.15..0.18$ for $H\approx H_L$ and then to $\beta_{\rm XY}
\simeq 0.346$ for $H>H_L$, in agreement with the experimental 
observation\cite{thermal.exp}.

From the low--temperature specific heat in the D phase 
one finds $E_g\approx 23\,\rm K$\cite{spec.heat,corr.lengths}, which
gives with Eq. (\ref{eq:C-phason})
for the phason specific heat $\beta_{\rm phason}\approx 
1.3\,\frac{\rm mJ}{\rm K^4mol}$ in the I phase, which has to be compared with
the experimental value of $1.4\,\frac{\rm mJ}{\rm K^4mol}$\cite{spec.heat}. This
good agreement is possibly to some degree
accidental, since the magnetic field dependences of the
various parameters have not been taken into account carefully. But at
least the order of magnitude should be right.

For the Larkin length we obtain with $x=n_{\rm imp}v_{\rm uc}/2$ for the concentration
of the Zn-atoms ($v_{\rm uc}$ denotes the volume per unit cell) and assuming 
a linear dependence of $T_{\rm SP}(x)$ on $x$ with  $d\ln{T_{\rm SP}(x)}/dx\approx 14$
\cite{khomskii}, for $x\approx 0.04 : L_L\approx 1.2\,\rm nm$ and 
for  $x\approx 0.07 : L_L\approx 0.7\,\rm nm$. 
The data of Kiryukhin et al. \cite{solitons.exp} was fitted with an
exponential decay of correlations with an anisotropic correlation
length $\xi $ of order $10\,{\rm nm}$. It would be interesting to
check, whether their data can also be fitted by a power law (\ref{eq:PP}).

The authors acknowledge discussions with M. Braden, S. Brasovskii, B. B\"uchner,
A. Gernoth, H. Schulz and G. Uhrig as well as the support from SFB 341
and GIF (T.N.). One of the authors (T.N.) acknowledges the hospitality of
ENS (Paris), where some of the work was done.

\end{multicols}

\begin{references}
\bibitem{fermi} Here, we consider the case in which the magnetic
  field decreases the Fermi level. The extension to the opposite case
  is straightforward.
\bibitem{foot} Although (\ref{eq:C-phason}) was derived from the
 assumption, that
$\psi$ obeys Bose statistics, dimensional arguments and the existence 
of a phason branch in the I phase lead to the same result, e.g. if we 
assume the dynamics of an Ising model in a transverse field \cite{struc.trans}.
\bibitem{deltac}The effect of randomness in the coefficients of the gradient 
terms is similar to that in $\delta r$ and follows from the above expressions,
if we replace $\delta r$ by $\delta c\,q_s^2$.
\end{references}
\end{document}